# A Saw-tooth Scaling of Work Function in Hydrogenated Graphene

Min Wang, Yan Zhou, Sui Kong Hark , Xi Zhu

School of Science and Engineering,

The Chinese University of Hong Kong, Shenzhen,

Shenzhen, Guangdong, 518172

ABSTRACT: The saw-tooth scaling investigations of work functions (WF) in hydrogenated graphene are theoretically performed by density functional theory. The positions of hydrogen atoms and the H/C ratios affect WF. With the comparison of different graphene structures by adsorbed two hydrogen atoms, WF is larger with less negative charges of the graphene. With different H/C ratios, WF is controlled by two parts: the graphene surface and hydrogen groups. When the ratio is smaller than 0.5, the work function is more dependent on the Mulliken charge of graphene. With more hydrogen atoms, the surface of graphene is more negative, resulting in a smaller work function. When the ratio is larger than 0.5, the effect of hydrogen groups cannot be ignored. When two sides of graphene have different numbers of hydrogen atoms, a dipole exists, making the movement of electron from Fermi level to vacuum level more difficult, thus WF is larger than that with the same hydrogen atoms of each side. Therefore the scaling of work function acts like a saw-tooth dependent on H/C ratio.

Since graphene is obviously observed in 2004[1], it has attracted tremendous interests of this new two-dimensional structure in nanotechnology and nanoscience[2]. Graphene has unique properties, including quantum hall effect[3], intrinsic mobility[4], and massless Dirac ferminons[4]. Its derivatives, like graphene oxide[5-11], graphene nanoribbons[12-19], hydrogenated graphene[20-22], and composite structures[23-27], also provide atomic understandings for the extension of graphene materials for various applications[28].

Graphene-based materials are ideal for flexible photonics[29-31] and electrodes[32-33], due to the optical transparent, sheet-like and conductive properties. In most electronics, work function (WF) is an essential parameter to evaluate the performance of the devices and to point out the way for the improvement as well. For example, to fabricate a device with an Ohmic contact of graphene and other layers rather than Schottky one could improve its energy conversion efficiency[34-36].

WF is always defined as the minimum energy to move an electron away from the Fermi level to the vacuum level[37] and it is also one of the parameters which can be investigated and proven by experimental and theoretical ways. In experiment, WF can be measured by Kelvin probe force microscopy[38]. In theory, WF can be obtained from the calculated potential distribution, which has been widely used for the investigations in many materials, including metals[39-40], carbon nanotubes[41-43], and BN sheets[44-45].

Here we select several hydrogenated graphene structures to investigate the functionalization dependent WF since with different experimental synthesis conditions[46-47], the hydrogen atoms on the graphene surface can also be controllable by patterned adsorption, and the hydrogenation near room temperature is achieved easier on monolayer graphene than that on bilayer graphene[48]. In this work, we investigate WF of hydrogenated graphene modulated by different H/C ratios to extend graphene-based materials' potentials in the electronics.

The theoretical density functional theory (DFT) calculations are performed with DMol3[49]. Local density approximation (LDA)[50] and Perdew-Wang (PWC) functional[51] are used for exchange and correlation, and the double numeric basis with d-functions (DND) is used as the atomic basis set. The tolerance of density convergence in self-consistent field (SCF), energy change, force, and displacement are $1 \times 10^{-5}$ a.u., $2 \times 10^{-5}$ Ha, 0.004 Ha/Å, and 0.005 Å, respectively. We choose the $3 \times 3$ graphene supercell as the based system. Then the structures of different numbers of hydrogen atoms connected with graphene are investigated. The chosen structures are schematically shown in Figure 2. All these structures are fully optimized. $11 \times 11 \times 1$ is used for the Monkhorst-Pack k-point grid to sample Brillouin zone. We use a 25 Å vacuum space along z axis where WF's value is calculated.

Since hydrogen atoms can not only connect with different carbon atoms but also both sides of graphene, we only investigate several structures, which gain low energy with the same ratios of hydrogen/carbon. To determine the strategies for the chosen structures, we first investigate several structures of two hydrogen atoms on the surfaces in Figure 1. The structures A-E with two hydrogen groups on both sides are shown in Figure 1 (a)-(e) respectively, while structure F of the hydrogen atoms connecting with the same carbon atoms as structure A but on one side of graphene is illustrated in (f). The total energy of each structure is calculated in Table 1 and the total of structure A is set to zero as the benchmark. Compared with structures A to E, A has the lowest energy and is most stable, revealing that two hydrogen atoms prefer to connect with two neighbored carbon atoms. With the consideration of A and F, the structure with two hydrogen groups connecting with the same carbon atoms on both sides of graphene is more stable. Different connections of two hydrogen groups can affect the values of WF (Table 1). It is also noticed that the Mulliken charges of graphene surfaces are also influenced by different adsorption positions (Table 1). The structure with more negative Mulliken charges of graphene has a smaller WF, since the movement of an electron is easier from the more negative structure. The Mulliken charge of graphene play an important role in the value of WF.

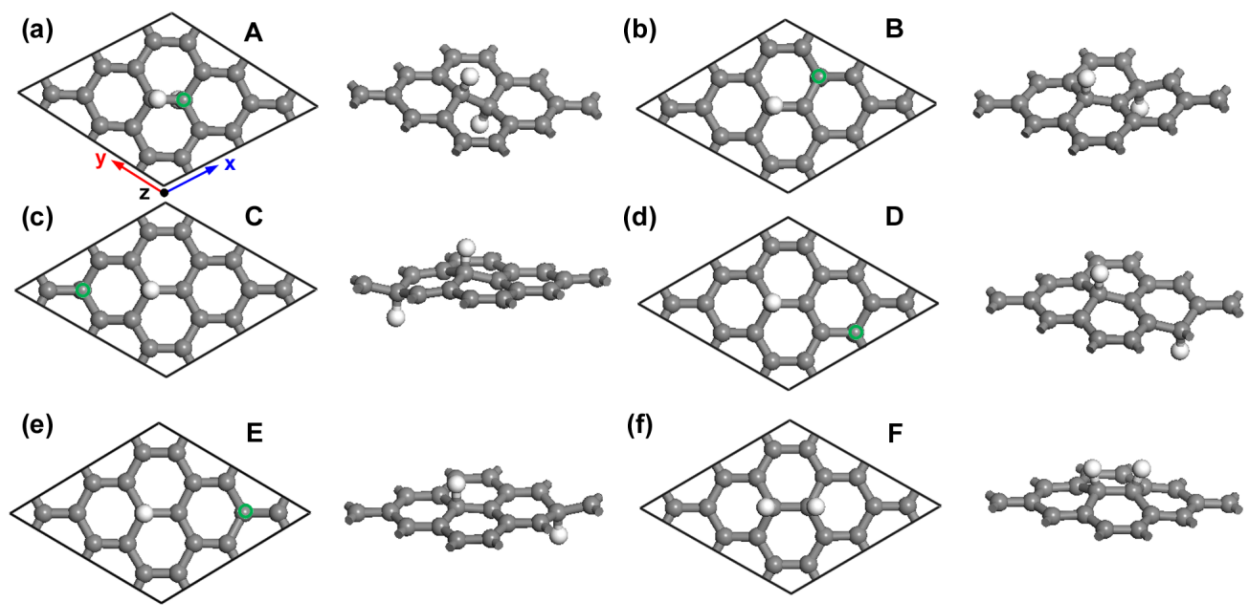

Figure 1 (a)-(f) The structures A-F of two hydrogen atoms connecting with different positions of graphene in the top- and side-view. Carbon and hydrogen atoms are colored in gray and white, respectively. The values are denoted as the total energy of each structure and that of structure A is set to zero. The green circles display the unseen connections between carbon atoms and hydrogen atoms on the back side.

Table 1. The energies, work functions (WF) and Mulliken charges of graphene in structures A to F. The units of energy and work function are $eV$ and that of charge is $e$.

| Structure | E-E$_A$ (eV) | WF (eV) | Mulliken charge (e) |
| --- | --- | --- | --- |
| A | 0 | 4.14 | -0.427 |
| B | 1.98 | 3.97 | -0.524 |
| C | 0.73 | 4.05 | -0.498 |
| D | 1.23 | 4.05 | -0.488 |
| E | 2.16 | 3.92 | -0.551 |
| F | 0.71 | 4.05 | -0.482 |

Then we study the structures with different H/C ratios to find the trend of WF. As mentioned above, the WF are dependent on the hydrogen positions on graphene surface and we cannot investigate all possible structures, so we only choose some more stable structures to study. Based on the discussion above, the choice strategies are listed: 1) hydrogen groups on both sides of graphene; 2) hydrogen groups connect with the neighbored carbon atoms. Therefore, for the structures with even numbers of hydrogen atoms, each side of graphene combines with the same number of hydrogen groups, but the structures with odd numbers of hydrogen atoms possess one side gaining more hydrogen groups. The studied 18 structures with neighbored connections on both sides of graphene are chosen and schematically shown in Figure 2.

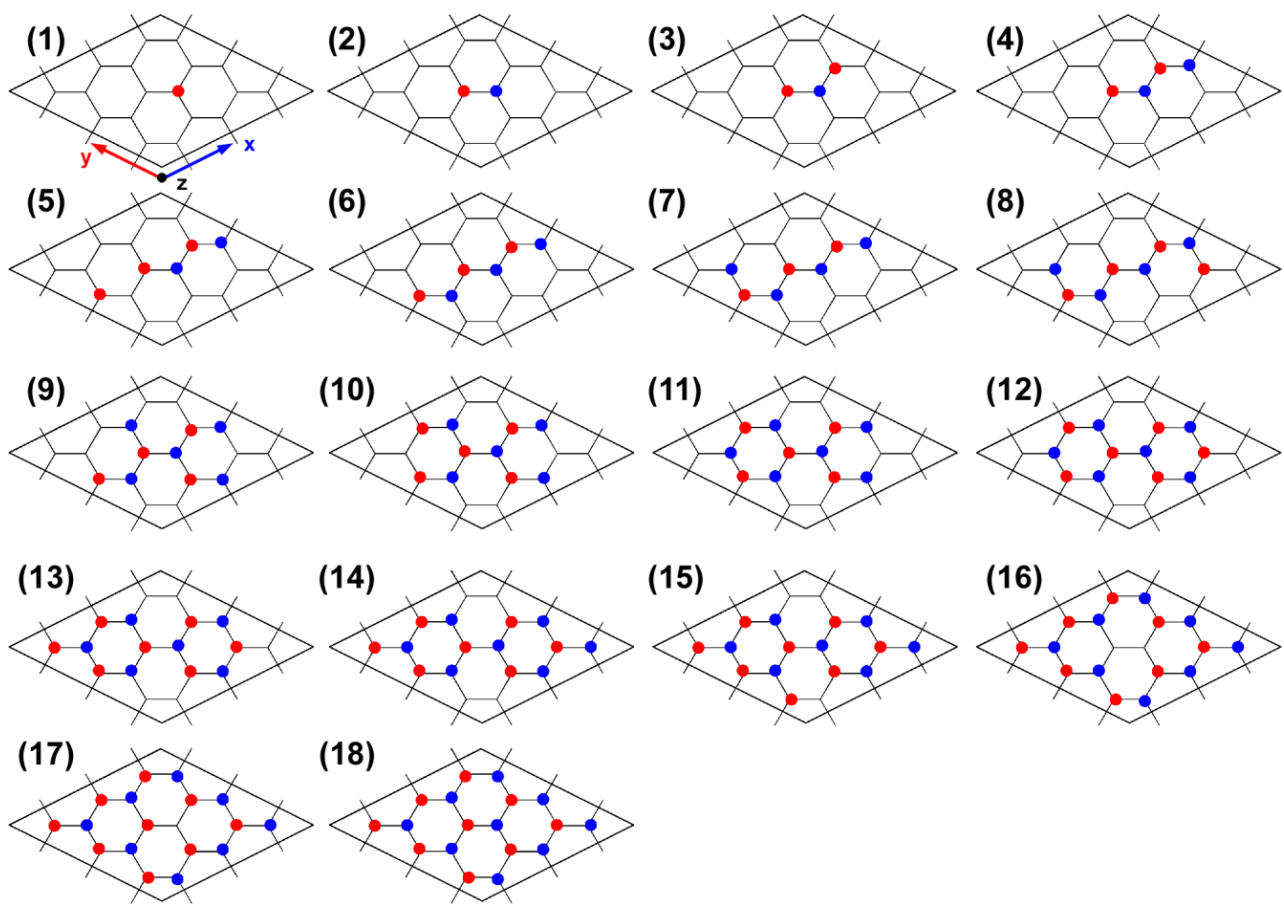

Figure 2. The hydrogenated graphene structures. The positions of hydrogen atoms connecting with the carbon atoms are denoted by red or blue dots to show the connection on the front or back side of graphene.

With adsorption of more hydrogen groups on the surface, the graphene becomes more negative, confirmed by the Mulliken charge in Table 2 and Figure 3, thus it can be considered that the trend of WF would be smaller with more hydrogen atoms. However, the calculated WF of the structures with different H/C ratios (see Figure 4) is a little more complicated to explain the origination only from Mulliken charge. But it is also clearly shown that WF is not only dependent on the hydrogen positions, but also has a relation with H/C ratio. When H/C ratio is smaller than 0.5, WF mainly decreases with the increase of the ratio. However, when the ratio is larger than 0.5, the variation becomes severer. The whole shape of WF dependent on H/C ratio is like a saw-tooth scaling. If we only see the structures with odd or even hydrogen atoms, it seems that the WF trend of these structures (odd or even) decrease with the increase of H/C ratio. Then the question about the effects of odd or even hydrogen atoms is coming consequently.

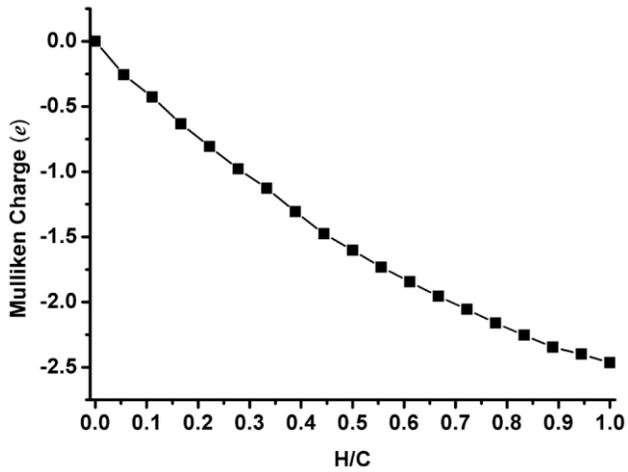

Figure 3. Mulliken charge is a function of H/C ratio.

Table 2. The H/C ratio, work function (WF) and Mulliken charge of graphene in the bare graphene and hydrogenated graphene structures (shown in Figure 2).

| Structure | H/C | WF (eV) | Mulliken charge (e) |
|---|---|---|---|
| graphene | 0.00 | 4.46 | 0.000 |
| 1 | 0.06 | 4.16 | -0.257 |
| 2 | 0.11 | 4.14 | -0.427 |
| 3 | 0.17 | 3.94 | -0.633 |
| 4 | 0.22 | 3.92 | -0.807 |
| 5 | 0.28 | 3.84 | -0.979 |
| 6 | 0.33 | 3.86 | -1.128 |
| 7 | 0.39 | 3.67 | -1.308 |
| 8 | 0.44 | 3.32 | -1.476 |
| 9 | 0.50 | 3.51 | -1.603 |
| 10 | 0.56 | 2.83 | -1.732 |

| 11 | 0.61 | 3.35 | -1.845 |
| 12 | 0.67 | 1.74 | -1.956 |
| 13 | 0.72 | 3.1  | -2.056 |
| 14 | 0.78 | 1.14 | -2.161 |
| 15 | 0.83 | 2.83 | -2.255 |
| 16 | 0.89 | 1.01 | -2.346 |
| 17 | 0.94 | 2.78 | -2.399 |
| 18 | 1.00 | 0.62 | -2.466 |

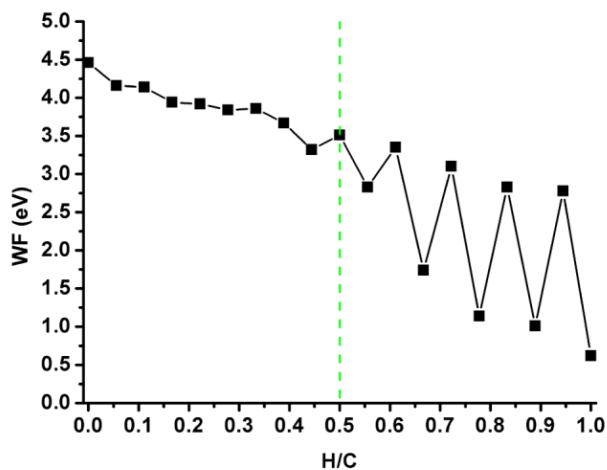

Figure 4 Work function is dependent on the H/C ratio.

To answer this question, the electrostatic potentials of structures 17 and 18 are plotted in Figure 5. The vacuum level is set to zero and denoted as a green dashed line. Each carbon atom of structure 18 connects with one hydrogen atom, while only one carbon atom of structure 17 does not connected with one hydrogen atom. Two curves of electrostatic potentials are similar due to their similar structures. The electrostatic potential around the middle of fractional coordinate in structure 17 is deeper than that in structure 18. Fermi level of structure 17 (the cyan dashed line) is also lower than that of structure 18 (the pink dashed line). Thus WF of structure 17 is larger than that of structure 18.

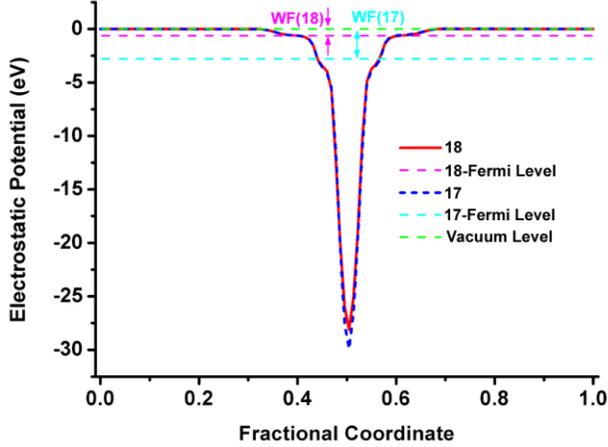

Figure 5 Electrostatic potentials of structures 17 and 18, denoted as blue dashed and red lines. Vacuum level is set to zero and displayed as a green dashed line. Fermi levels for structures 17 and 18 are denoted as cyan and pink dashed lines. WF for two structures are also marked in cyan and pink for structures 17 and 18 respectively.

Moreover, with odd hydrogen atoms, one side of graphene gains more hydrogen atoms, thus it can be considered that an existence of a dipole along two sides, while the structure with even hydrogen atoms, two sides of graphene gains the same number of hydrogen groups, resulting in no dipole. To figure out this dipole phenomenon, 2D deformation densities of structure 17 and 18 are plotted in Figure 6 (a) and (b) respectively. Deformation density[52] is defined as the difference between the structure's charge density and the free atom's charge density. Then the side with more hydrogen atoms is more positive, making the movement of electron from Fermi level to vacuum level more difficult, therefore WF is larger than that with even hydrogen atoms nearby (see Figure 5). When H/C ratio is smaller than 0.5, graphene plays an important role in WF, but when H/C ratio is larger than 0.5, the effects of the dipole caused by odd hydrogen atoms cannot be ignored and it is the reason for the saw-tooth scaling of WF.

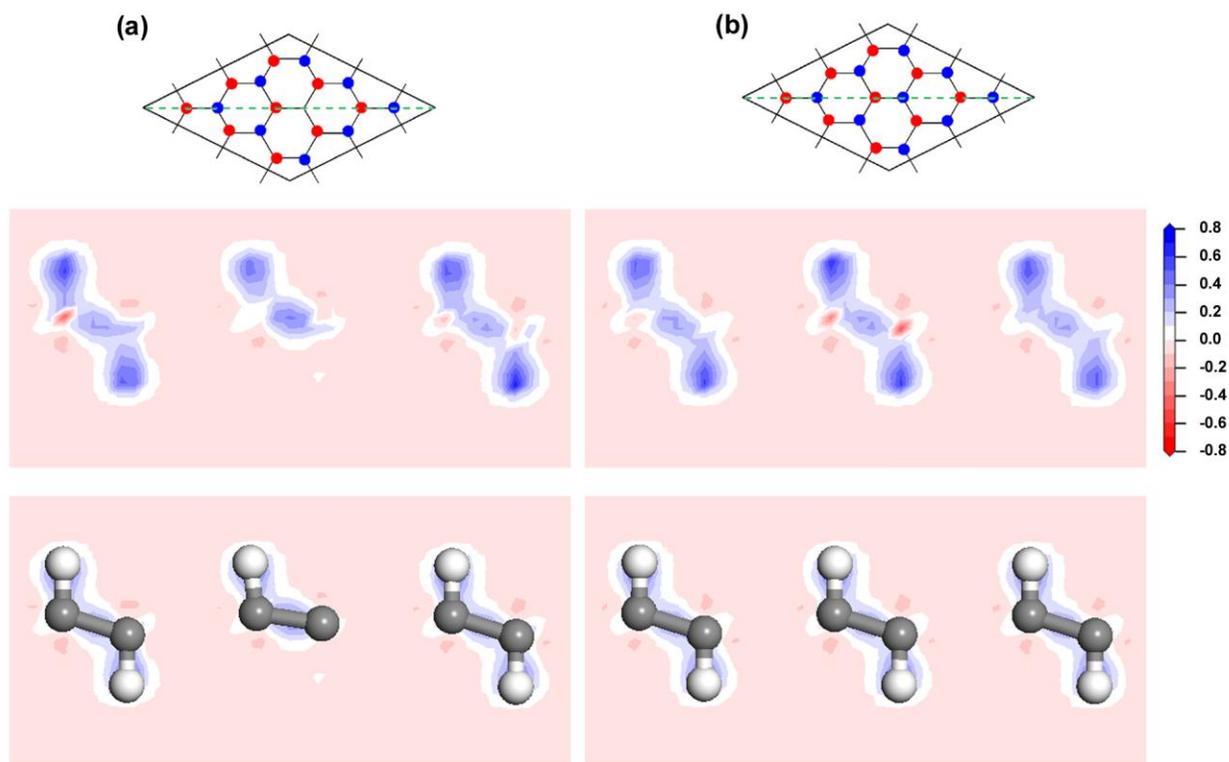

Figure 6. 2D deformation density for structure (a) 17 and (b) 18. The green dashed lines in three structures are used to point out the mapping slices for deformation density. The legend of color map is shown on the right. The white and grey balls are denoted as hydrogen and carbon atoms.

In conclusion, the saw-tooth scaling investigations of work functions in hydrogenated graphene are theoretically performed by density functional theory. The positions of hydrogen atoms and the H/C ratios affect WF. With the comparison of different graphene structures by adsorbed two hydrogen atoms, WF is larger with less negative Mulliken charges of the graphene. With different H/C ratios, WF is controlled by two parts: the graphene and hydrogen groups. When the ratio is smaller than 0.5, WF is more dependent on the Mulliken charge of graphene. With more hydrogen atoms, the surface of graphene is more negative, resulting in a smaller work function. When the ratio is larger than 0.5, the effect of the hydrogen atoms cannot be ignored. When two sides of graphene have different numbers of hydrogen atoms, a dipole exists and then the side with more hydrogen is more positive, making the movement of electron from Fermi level to vacuum level more difficult, thus the work function is larger than that with the same hydrogen atoms of each side. Therefore the scaling of work function acts like a saw-tooth dependent on H/C ratio.

**Acknowledgment**

This work is partially financially supported by

**REFERENCES**

1. Novoselov, K. S.; Geim, A. K.; Morozov, S. V.; Jiang, D.; Zhang, Y.; Dubonos, S. V.; Grigorieva, I. V.; Firsov, A. A., Science 2004, 306, 666-669.
2. Castro Neto, A. H.; Guinea, F.; Peres, N. M. R.; Novoselov, K. S.; Geim, A. K., Reviews of Modern Physics 2009, 81, 109.
3. Zhang, Y. B.; Tan, Y. W.; Stormer, H. L.; Kim, P., Nature 2005, 438, 201-204.
4. Novoselov, K. S.; Geim, A. K.; Morozov, S. V.; Jiang, D.; Katsnelson, M. I.; Grigorieva, I. V.; Dubonos, S. V.; Firsov, A. A., Nature 2005, 438, 197-200.
5. Xu, Y. X.; Bai, H.; Lu, G. W.; Li, C.; Shi, G. Q., Journal of the American Chemical Society 2008, 130, 5856-+.
6. Jeong, H. K.; Lee, Y. P.; Lahaye, R.; Park, M. H.; An, K. H.; Kim, I. J.; Yang, C. W.; Park, C. Y.; Ruoff, R. S.; Lee, Y. H., Journal of the American Chemical Society 2008, 130, 1362-1366.
7. Wang, M.; Li, C. M., New Journal of Physics 2010, 12, 083040.
8. Li, Z. Y.; Zhang, W. H.; Luo, Y.; Yang, J. L.; Hou, J. G., Journal of the American Chemical Society 2009, 131, 6320-+.
9. Wang, M.; Li, C. M., Physical Chemistry Chemical Physics 2011, 13, 1413-1418.
10. Liu, Z.; Robinson, J. T.; Sun, X. M.; Dai, H. J., Journal of the American Chemical Society 2008, 130, 10876-+.
11. Wang, M.; Huang, W.; Chan-Park, M. B.; Li, C. M., Nanotechnology 2011, 22, 105702.
12. Kan, E. J.; Li, Z. Y.; Yang, J. L.; Hou, J. G., Journal of the American Chemical Society 2008, 130, 4224-+.
13. Zhu, X.; Su, H. B., Journal of Physical Chemistry C 2010, 114, 17257-17262.
14. Son, Y. W.; Cohen, M. L.; Louie, S. G., Nature 2006, 444, 347-349.
15. Wang, M.; Li, C. M., Nanoscale 2012, 4, 1044-1050.
16. Son, Y. W.; Cohen, M. L.; Louie, S. G., Physical Review Letters 2006, 97, 216803.
17. Zhu, X.; Su, H. B., Journal of Physical Chemistry A 2011, 115, 11998-12003.

18. Radovic, L. R.; Bockrath, B., Journal of the American Chemical Society 2005, 127, 5917-5927.

19. Yang, X. Y.; Dou, X.; Rouhanipour, A.; Zhi, L. J.; Rader, H. J.; Mullen, K., Journal of the American Chemical Society 2008, 130, 4216-+.

20. Boukhvalov, D. W.; Katsnelson, M. I.; Lichtenstein, A. I., Physical Review B 2008, 77, 035427.

21. Wang, M.; Li, C. M., Nanoscale 2011, 3, 2324-2328.

22. Xie, L. M.; Jiao, L. Y.; Dai, H. J., Journal of the American Chemical Society 2010, 132, 14751-14753.

23. Wang, M.; Li, C. M., Nanotechnology 2010, 21, 035704.

24. Su, H. B.; Goddard, W. A.; Zhao, Y., Nanotechnology 2006, 17, 5691-5695.

25. Wang, M.; Li, C. M., Physical Chemistry Chemical Physics 2011, 13, 5945-5951.

26. Zhu, X.; Su, H. B., Physical Review B 2009, 79, 165401.

27. Su, H. B.; Nielsen, R. J.; van Duin, A. C. T.; Goddard, W. A., Physical Review B 2007, 75, 134107.

28. Geim, A. K.; Novoselov, K. S., Nature Materials 2007, 6, 183-191.

29. Hasan, T.; Sun, Z. P.; Wang, F. Q.; Bonaccorso, F.; Tan, P. H.; Rozhin, A. G.; Ferrari, A. C., Advanced Materials 2009, 21, 3874-3899.

30. Xia, F. N.; Mueller, T.; Lin, Y. M.; Valdes-Garcia, A.; Avouris, P., Nature Nanotechnology 2009, 4, 839-843.

31. Bonaccorso, F.; Sun, Z.; Hasan, T.; Ferrari, A. C., Nature Photonics 2010, 4, 611-622.

32. Kim, K. S.; Zhao, Y.; Jang, H.; Lee, S. Y.; Kim, J. M.; Ahn, J. H.; Kim, P.; Choi, J. Y.; Hong, B. H., Nature 2009, 457, 706-710.

33. Eda, G.; Fanchini, G.; Chhowalla, M., Nature Nanotechnology 2008, 3, 270-274.

34. Shi, Y. M.; Kim, K. K.; Reina, A.; Hofmann, M.; Li, L. J.; Kong, J., Acs Nano 2010, 4, 2689-2694.

35. Chandramohan, S.; Kang, J. H.; Katharria, Y. S.; Han, N.; Beak, Y. S.; Ko, K. B.; Park, J. B.; Kim, H. K.; Suh, E. K.; Hong, C. H., APPLIED PHYSICS LETTERS 2012, 100, 023502.

36. Huang, J. H.; Fang, J. H.; Liu, C. C.; Chu, C. W., Acs Nano 2011, 5, 6262-6271.

37. Lang, N. D.; Kohn, W., Physical Review B 1971, 3, 1215-1223.

38. Melitz, W.; Shen, J.; Kummel, A. C.; Lee, S., Surface Science Reports 2011, 66, 1-27.

39. Paggel, J. J.; Wei, C. M.; Chou, M. Y.; Luh, D. A.; Miller, T.; Chiang, T. C., Physical Review B 2002, 66, 233403.

40. Park, S.; Colombo, L.; Nishi, Y.; Cho, K., APPLIED PHYSICS LETTERS 2005, 86, 073118.

41. Shan, B.; Cho, K., Physical Review Letters 2005, 94, 236602.

42. Shan, B.; Cho, K., Physical Review B 2006, 73, 081401.

43. Su, W. S.; Leung, T. C.; Chan, C. T., Physical Review B 2007, 76, 235413.

44. Jiao, N.; He, C. Y.; Zhang, C. X.; Peng, X. Y.; Zhang, K. W.; Sun, L. Z., Aip Advances 2012, 2, 022125.

45. Xie, Y.; Yu, H. T.; Zhang, H. X.; Fu, H. G., Physical Chemistry Chemical Physics 2012, 14, 4391-4397.

46. Elias, D. C.; Nair, R. R.; Mohiuddin, T. M. G.; Morozov, S. V.; Blake, P.; Halsall, M. P.; Ferrari, A. C.; Boukhvalov, D. W.; Katsnelson, M. I.; Geim, A. K.; Novoselov, K. S., Science 2009, 323, 610-613.

47. Balog, R.; Jorgensen, B.; Nilsson, L.; Andersen, M.; Rienks, E.; Bianchi, M.; Fanetti, M.; Laegsgaard, E.; Baraldi, A.; Lizzit, S.; Sljivancanin, Z.; Besenbacher, F.; Hammer, B.; Pedersen, T. G.; Hofmann, P.; Hornekaer, L., Nature Materials 2010, 9, 315-319.

48. Ryu, S.; Han, M. Y.; Maultzsch, J.; Heinz, T. F.; Kim, P.; Steigerwald, M. L.; Brus, L. E., Nano Letters 2008, 8, 4597-4602.

49. Dmol3 is available as part of Materials Studio.Accelrys Inc.`, Suite 100 San Diego`, CA 92121`, USA.

50. Troullier, N.; Martins, J. L., Physical Review B 1991, 43, 1993-2006.

51. Perdew, J. P.; Wang, Y., Physical Review B 1992, 45, 13244-13249.

52. Hirshfeld, F. L., Theor. Chim. Acta B 1977, 44, 129.